\documentclass[sigconf]{acmart}

\usepackage{algorithm}
\usepackage{algorithmic}
\usepackage{array}
\usepackage{amsfonts}
\usepackage{amssymb}
\usepackage{amsmath}
\usepackage{bm}
\usepackage{booktabs}
\usepackage{color}
\usepackage{epstopdf}
\usepackage{enumitem}
\usepackage{footmisc}
\usepackage{graphicx}
\usepackage{hyperref}
\usepackage{listings}
\usepackage{multirow}
\usepackage{mathrsfs}
\usepackage{pifont}
\usepackage{subfigure}
\usepackage{setspace}
\usepackage{soul}
\usepackage{textcomp}
\usepackage{url}
\usepackage{xcolor}
\usepackage{xspace}
\usepackage[normalem]{ulem}
\usepackage[np, autolanguage]{numprint}
\usepackage[skip=0pt]{caption}
\usepackage{svg}

\newcommand{\paratitle}[1]{\vspace{1.5ex}\noindent\textbf{#1}}
\newcommand{\ie}{\emph{i.e.,}\xspace}

\newcommand{\eg}{\emph{e.g.,}\xspace}

\newcommand{\wrt}{w.r.t.\xspace}
\newcommand{\ignore}[1]{}

\newcommand{\modelname}{IDA-SR}

\begin{document}
\thanks{$^\dagger$ Corresponding author}
\author{Shanlei Mu$^{1}$, Yupeng Hou$^{2}$, Wayne Xin Zhao$^{2\dagger}$, Yaliang Li$^{3}$, and Bolin Ding$^{3}$}
\affiliation{
  $^1$School of Information, Renmin University of China, China \\
  $^2$Gaoling School of Artificial Intelligence, Renmin University of China, China \\
  $^3$Alibaba Group, USA \\
  \country{}
}
\email{slmu@ruc.edu.cn}
\fancyhead{}

\title[ID-Agnostic User Behavior Pre-training for Sequential Recommendation]{\texorpdfstring{ID-Agnostic User Behavior Pre-training for\\ Sequential Recommendation}{ID-Agnostic User Behavior Pre-training for Sequential Recommendation}}

\begin{abstract}
Recently, sequential recommendation has emerged as a widely studied topic.
Existing researches mainly design effective neural architectures to model user behavior sequences based on item IDs.
However, this kind of approach highly relies on user-item interaction data and neglects the attribute- or characteristic-level correlations among similar items preferred by a user.
In light of these issues, we propose \textbf{\modelname}, which stands for \textbf{ID}-\textbf{A}gnostic User Behavior Pre-training approach for \textbf{S}equential \textbf{R}ecommendation.
Instead of explicitly learning representations for item IDs, \modelname~directly learns item representations from rich text information.
To bridge the gap between text semantics and sequential user behaviors,
we utilize the pre-trained language model as text encoder, and conduct a pre-training architecture on the sequential user behaviors.
In this way, item text can be directly utilized for sequential recommendation without relying on item IDs.   
Extensive experiments show that the proposed approach can achieve comparable results when only using ID-agnostic item representations, and performs better than baselines by a large margin when fine-tuned with ID information.
\end{abstract}

\maketitle

\section{Introduction}
\label{sec-intro}
Over the past decade, sequential recommendation has been a widely studied task~\cite{sr-survey}, which learns time-varying user preference and provides timely information resources in need.
To better model sequential user behaviors, recent approaches~\cite{GRU4Rec, NARM, Caser, SASRec, BERT4Rec} mainly focus on how to design effective neural architecture for recommender system, such as Recurrent Neural Network (RNN)~\cite{GRU4Rec, NARM} and Transformer~\cite{SASRec, BERT4Rec}.
Typically, existing methods learn the item representations based on item IDs (\ie a unique integer associated with an item) and then feed them to the designed sequential recommender. 
We refer to such a paradigm as \emph{ID-based item representation}.

In the literature of recommender systems,  ID-based item representation has been a mainstream paradigm to model items and develop recommendation models. It is conceptually simple and flexibly extensible. 
However, it is also noticed with  several limitations for sequential recommendation. 
First, it highly relies on user-item interaction data to learn ID-based item representations~\cite{FPMC, GRU4Rec}. 
When interaction data is insufficient, it is difficult to derive high-quality item representations, which are widely observed in practice~\cite{ASRep, S3Rec}.
Second, there is a discrepancy between real user behaviors and the learned sequential models. 
Intuitively, user behaviors are driven by underlying user preference, \ie the preference over different item attributes or characteristics, so that a user behavior sequence essentially reflects the attribute- or characteristic-level correlations among similar items preferred by a user.
However, ID-based item representations learn  more abstractive ID-level correlations, which  cannot directly  characterize fine-grained correlations that actually reflect real user preference. 

Considering the above issues, we aim to represent items in a more natural way from the user view, such that it can directly capture attribute- or characteristic-level correlations without explicitly involving item IDs.
The basic idea is to incorporate item side information in modeling item sequences, where we learn \emph{ID-agnostic item representations} that are derived from attribute- or characteristic-level correlations driven by user preference. 
Specifically, in this work, we utilize rich text information to represent items instead of learning ID-based item representations.
As a kind of natural language,  item text reflects the human's cognition about item attributes or characteristics, which provides a general information resource to reveal user preference from sequential behavior. 
Different from existing studies that leverage item text to enhance ID-based representations~\cite{TCF,FDSA} or improve zero-shot recommendations~\cite{ZESRec}, this work aims to learn sequential user preference solely based on text semantics without explicitly modeling item IDs.

However, it is not easy to transfer the semantics reflected in the item texts to the recommendation task, as the semantics may not be directly relevant to the user preference or even noisy to the recommendation task~\cite{NARRE}. 
There is a natural semantic gap between natural language and user behaviors.
To fill this gap, we construct a pre-trained user behavior model that learns user preference by modeling the text-level correlations through behavior sequences.

To this end, we propose an \textbf{ID}-\textbf{A}gnostic User Behavior Pre-training approach for \textbf{S}equential \textbf{R}ecommendation , named \textbf{\modelname}.
Compared with existing sequential recommenders, the most prominent feature of \modelname~is that it no longer explicitly learn representations for item IDs.  Instead, it directly learns item representations from item texts, and we utilize  
the pre-trained language model~\cite{BERT, GPT3} as text encoder to represent items.
To adapt text semantics to the recommendation task, the key point is a pre-training architecture conducted on the sequential user behaviors with three important pre-training tasks, namely  \emph{next item prediction}, \emph{masked item prediction} and \emph{permuted item prediction}.
In this way, our approach can effectively bridge the gap between text semantics and sequential user behaviors, so that item text can be directly utilized for sequential recommendation without the help of item IDs.   

To evaluate the proposed approach, we conduct extensive experiments on the four real-world datasets.
Experimental results demonstrate that only using ID-agnostic item representations, the proposed approach can achieve comparable results with several competitive recommendation methods and even much better when training data is limited.
In order to better fit the recommendation task, we also provide a fine-tuning mechanism that allows the pre-trained architecture to use the guidance of explicit item ID information.
In such way, our approach performs better than baseline methods by a large margin under various settings, which is brought by the benefit of the ID-agnostic user behavior pre-training.

\section{Methodology}
\label{sec-model}
\begin{figure}[htb]
    \centering\includegraphics[width=0.47\textwidth]{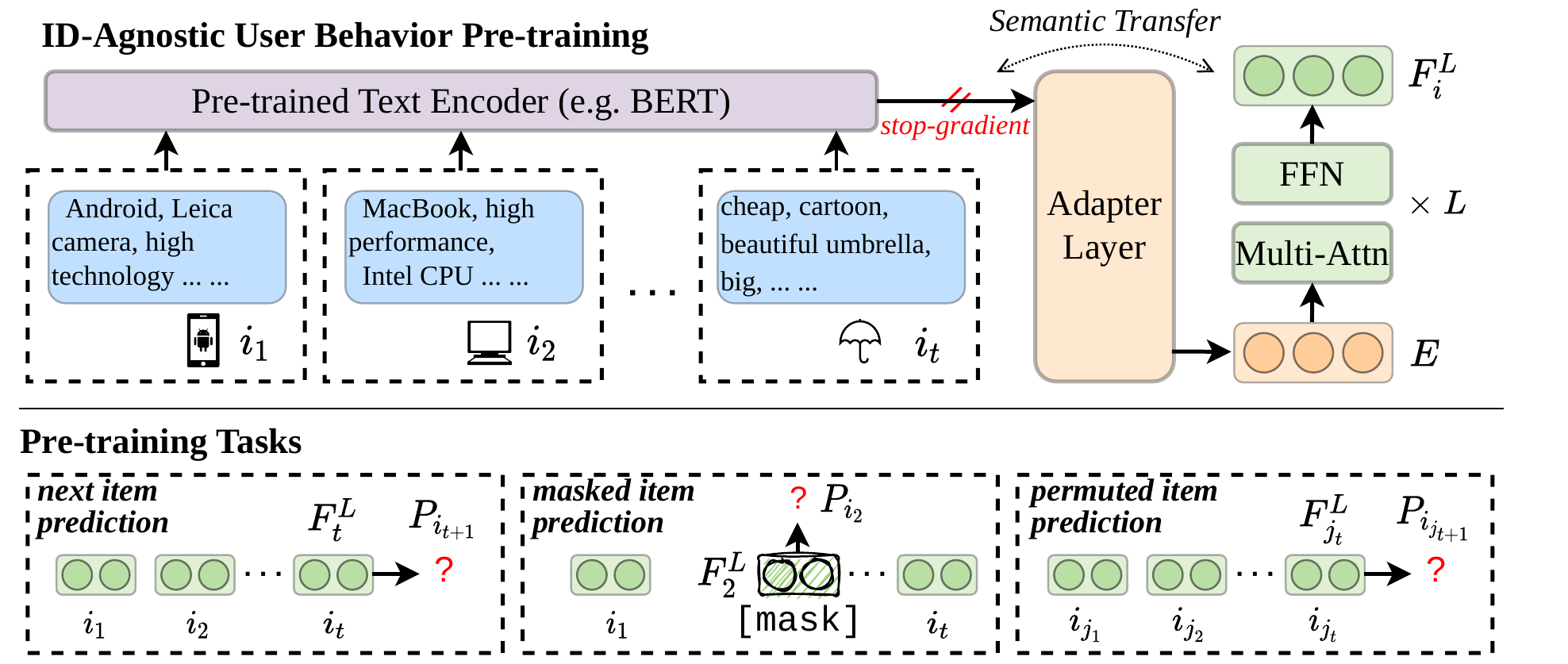}
    \caption{Overall illustration of the proposed approach.}
    \vspace{-0.15in}
    \label{fig-model}
\end{figure}

Given the ordered sequence of user $u$'s historical items up to the timestamp-$t$: $\{i_{1}, \cdots, i_{t}\}$, we need to predict the next item, \ie $i_{t+1}$.
In this section, we present the \textbf{ID}-\textbf{A}gnostic user behavior modeling approach for \textbf{S}equential \textbf{R}ecommendation, named \textbf{\modelname}.
It consists of \emph{ID-agnostic user behavior pre-training stage} and \emph{fine-tuning stage}.
The major novelty lies in the ID-agnostic user behavior pre-training, where we represent items solely based on item texts instead of item IDs. 
To adapt text semantics to the recommendation task, we incorporate an adapter layer that transforms text representations into item representations, and further design three elaborate pre-training tasks to retain the preference characteristics based on user behavior sequences. 
Figure~\ref{fig-model} presents the overview of the pre-training stage.
Next, we describe our approach in detail.

\subsection{ID-Agnostic User Behavior Pre-training}
For sequential recommendation, the key point of learning ID-agnostic representations is to capture 
sequential preference characteristics from user behavior sequences based on item texts. 
Since item texts directly describe items' attributes or characteristics, our pre-training approach tries to integrate item  characteristic encoding  into sequential user behavior modeling, and further bridge the semantic gap between them. 

\subsubsection{ID-Agnostic Item Representations}
\label{subsubsec-id}
To obtain ID-agnostic item representations, our idea is to utilize the pre-trained language model~(PLM)~\cite{BERT, GPT3}  to encode item texts.
Specifically, we use the pre-trained language model BERT as the text encoder to generate item representations based on the item text.
Given an item $i$ and its corresponding item text $C_{i}=\{w_{1},\cdots,w_{m}\}$, we first add an extra token {\tt [CLS]} into the item text $C_{i}$ to form the input word sequence $\tilde{C}_{i} = \{{\tt [CLS]}, w_{1},\cdots,w_{m}\}$.
Then the input word sequence $\tilde{C}_{i}$ is fed to the pre-trained BERT model.
We use the embedding for the {\tt [CLS]} token as ID-agnostic item representations.
In this way, we can obtain the item text embedding matrix $\mathbf{M}_{T}\in\mathbb{R}^{|\mathcal{I}| \times \tilde{d}}$, where $\tilde{d}$ is the dimension of item text embedding.

Different from ID-based item representations,  ID-agnostic item representations are less sensitive to the quality of the interaction data. Instead, it allows the sequential model to capture attribute- characteristics preference from user behavior sequences. It is also more resistible to the cold-start scenarios where a new item occurs for recommendation. 

\subsubsection{Text Semantic Adapter Layer}
Although the ID-agnostic item representations generated from PLMs have great expressive ability for item characteristics, not all the encoded semantics in these representations are directly beneficial or useful for sequential user behavior modeling. Therefore, we incorporate a text semantic adapter layer for transforming the original text representations into a form that is more suitable for the recommendation task.
The adapter layer is formalized as follows:
\begin{equation}
    \tilde{\bm{m}}_{i} = \sigma\big(\sigma(\bm{m}_{i}\mathbf{W}_{1} + \bm{b}_{1})\mathbf{W}_{2} + \bm{b}_{2}\big),
\end{equation}
where $\bm{m}_{i} \in \mathbf{M}_{T}$ is the input item representation, $\tilde{\bm{m}}_{i} \in \mathbb{R}^{1 \times d}$ is the updated item representation, $\mathbf{W}_{1} \in \mathbb{R}^{\tilde{d} \times d}$, $\mathbf{W}_{2} \in \mathbb{R}^{d \times d}$ and  $\bm{b}_{1},\bm{b}_{2} \in \mathbb{R}^{1 \times d}$ are learnable parameters, $\sigma(\cdot)$ is the activation function.
So we can obtain the updated item text embedding matrix $\tilde{\mathbf{M}}_{T} \in \mathbb{R}^{|\mathcal{I}| \times d}$.

Given a $n$-length item sequence, we apply a look-up operation from $\tilde{\mathbf{M}}_{T}$ to form the input embedding matrix $\mathbf{E}_{T} \in \mathbb{R}^{n \times d}$.
Besides, following \cite{SASRec}, we further incorporate a learnable position encoding matrix $\mathbf{P} \in \mathbb{R}^{n \times d}$ to enhance the input representation of the item sequence.
By this means, the sequence representations $\mathbf{E} \in \mathbb{R}^{n \times d}$ can be obtained by summing two embedding matrices $\mathbf{E} = \mathbf{E}_{T} + \mathbf{P}$.

\subsubsection{Sequential User Behavior Modeling}
The core of sequential recommendation lies in the sequential user behavior modeling, where we aim to capture sequential preference characteristics from user behavior sequences. 
Here, we adopt a classic self-attention architecture~\cite{Trans} based on the above ID-agnostic item representations.
A self-attention block generally consists of two sub-layers, \ie a multi-head self-attention layer (denoted by MultiHeadAttn($\cdot$)) and a point-wise feed-forward network (denoted by FFN($\cdot$)).
The update process can be formalized as following:
\begin{equation}
    \mathbf{F}^{l+1} = \text{FFN}(\text{MultiHeadAttn}(\mathbf{F}^{l})),
\end{equation}
where the $\mathbf{F}^{l}$ is the $l$-th layer's input.
When $l=0$, we set $\mathbf{F}^{0}=\mathbf{E}$.

\subsubsection{Text Semantics based User Behavior Pre-training Task}
Given the ID-agnostic item representations and  self-attention architecture, we next study 
how to design suitable optimization objectives to learn the parameters of the architecture, which is the key to bridge the semantic gap between text semantics and sequential preference characteristics. Next, we introduce three pre-training tasks derived from user behavior sequences based on text representations. 

\paratitle{Next Item Prediction.}
The \emph{next item prediction} pre-training task refers to predicting the next item having read all the previous ones, which has been widely adopted in the existing sequential recommendation methods~\cite{SASRec}.
Based on this task, we calculate the user's preference over the candidate item set as follows:
\begin{equation}
\label{eq-ar}
    P_{pre}(i_{t+1}|S) = {\text{softmax}(\mathbf{F}_{t}^{L}\tilde{\mathbf{M}}^{\top}_{T})}_{[i_{t+1}]},
\end{equation}
where $S = i_{1:t}$ is the user historical sequence, $\mathbf{F}_{t}^{L}$ is the output of the $L$-layer self-attention block at step $t$.
Such a task tries to capture sequential preference based on text semantics. 

\paratitle{Masked Item Prediction.}
The \emph{masked item prediction} pre-training task refers that corrupting the input item sequence and trying to reconstruct the original item sequence.
Specifically, we randomly mask some items (\ie replace them with a special token {\tt [MASK]}) in the input sequences, and then predict the masked items based on their surrounding context.
Based on this task, we calculate the user's preference over the candidate item set as follows:
\begin{equation}
\label{eq-ae}
    P_{pre}(i_{t}|S) = {\text{softmax}(\mathbf{F}_{t}^{L}\tilde{\mathbf{M}}^{\top}_{T})}_{[i_{t}]},
\end{equation}
where $S$ is the masked version for user behavior sequence, position $t$ is replaced with {\tt [MASK]}.
Such a task enhances the overall sequential modeling capacity of the recommendation model. 

\paratitle{Permuted Item Prediction.}
The \emph{permuted item prediction} pre-training task refers that permuting the items in the original user behavior sequence, then using the previous items in permuted user behavior sequence to predict the next item.
Given the input user behavior sequence $i_{1:t}=\{i_{1},\cdots,i_{t}\}$, we first generate the permuted user behavior sequence $i_{j_{1}:j_{t}}=\{i_{j_{1}},\cdots,i_{j_{t}}\}$ by permuting the items.
Then, based on the permuted sequence, we calculate the user's preference over the candidate item set as follows:
\begin{equation}
\label{eq-bar}
    P_{pre}(i_{j_{t+1}}|S) = {\text{softmax}(\mathbf{F}_{j_{t}}^{L}\tilde{\mathbf{M}}^{\top}_{T})}_{[i_{j_{t+1}}]},
\end{equation}
where $S=i_{j_{1}:j_{t}}$ is the permuted user historical sequence.
In this way, the context for each position consist of tokens from both left and right, which is able to improve the performance~\cite{XLNet}.

To combine the three pre-training tasks, 
we adopt the cross-entropy loss to pre-train our model as follows:
\begin{equation}
    \mathcal{L}_{pre} = - \sum_{u \in \mathcal{U}}\sum_{t \in \mathcal{T}}\log P_{pre}(i_{t}=i_{t}^{*}|S),
\end{equation}
where $i_{t}^{*}$ is the ground truth item, $\mathcal{T}$ is the predicted position set.

\subsection{Fine-tuning for Recommendation}
At the pre-training stage, we integrate the text semantics of items into the sequential behavior modeling. 
Next, we further optimize the architecture according to the recommendation task. 
Different from previous self-supervised recommendation models~\cite{BERT4Rec, S3Rec}, we  can fine-tune our approach \emph{with} or \emph{without} item IDs. 

\paratitle{Fine-tuning without ID.}
Without adding any extra parameters, we can directly fine-tune the pre-trained model based on the ID-agnostic item representations.
In this way, we calculate the user's preference score for the item $i$ in the step $t$ under the context from user history as:
\begin{equation}
\label{eq-fine}
    P_{fine}(i_{t+1}|i_{1:t}) = {\text{softmax}(\mathbf{F}_{t}^{L}\tilde{\mathbf{M}}_{T}^{\top})}_{[i_{t+1}]},
\end{equation}
where $\tilde{\mathbf{M}}_{T}$ is the updated item text embedding matrix, $\mathbf{F}_{t}^{L}$ is the output of the $L$-layer self-attention block at step $t$.
Since no additional parameters are incorporated, it enforces the model to well fit the recommendation task in an efficient way. 

\paratitle{Fine-tuning with ID.}
Unlike text information,  IDs are more discriminative to represent an item, \eg  an item will be easily identified when we know its ID. Therefore, we can  further improve the  discriminative ability of the above pre-trained approach by incorporating additional item ID representations.
Specifically, we maintain a learnable item embedding matrix $\mathbf{M}_{I} \in \mathbb{R}^{|\mathcal{I}|\times d}$.
Then we combine $\mathbf{M}_{I}$ and the item text embedding $\tilde{\mathbf{M}}_{T}$ as the final item representation.
We calculate the user's preference score for the item $i$ in the step $t$ under the context from user history as:
\begin{equation}
\label{eq-fine-id}
    P_{fine}(i_{t+1}|i_{1:t}) = {\text{softmax}(\tilde{\mathbf{F}}_{t}^{L}\tilde{(\mathbf{M}}_{T} + \mathbf{M}_{I})^{\top})}_{[i_{t+1}]}.
\end{equation}
Here,  we only incorporate item IDs at the fine-tuning stage and the rest parts (ID-agnostic item representations, adapter layer, and self-attention architecture) have been pre-trained at the pre-training stage. 
As will be shown in Section~\ref{sec-main-exp}, this fine-tuning method is more effective than that simply combining text and ID features.

For each setting, we adopt the widely used cross-entropy loss to train the model in the fine-tuning stage.

\section{EXPERIMENTS}
\label{sec-exp}
\subsection{Experimental Setup}

\begin{table}[htbp]
\vspace{-0.1in}
\centering
\caption{Statistics of the datasets.}
\label{tab-data}
\setlength{\tabcolsep}{0.5mm}{
	\begin{tabular}{l
	p{1.5cm}<{\centering} p{1.5cm}<{\centering} 
	p{1.5cm}<{\centering} p{1.5cm}<{\centering}}
    \toprule
    Dataset & Pantry & Instruments & Arts & Food \\
    \midrule
    \# Users & 13,101 & 24,962 & 45,486 & 115,349 \\
    \# Items & 4,898 & 9,964 & 21,019 & 39,670 \\
    \# Actions & 126,962 & 208,926 & 395,150 & 1,027,413 \\
    \bottomrule
    \end{tabular}}
\vspace{-0.1in}
\end{table}

\begin{table*}[t!]
    \small
    \caption{Performance comparison of different methods on the four datasets. The best performance and the best performance baseline are denoted in bold and underlined fonts respectively.
    % \emph{Improv.} indicates the relative improvements of the proposed approach over the best performance baselines.
    }
	\label{tab-main-exp}
	\resizebox{2.1\columnwidth}{!}{
	\setlength{\tabcolsep}{1mm}{
	\begin{tabular}{ll
	p{1.1cm}<{\centering} p{1.1cm}<{\centering} p{1.1cm}<{\centering}
	p{1.1cm}<{\centering} p{1.1cm}<{\centering} p{1.1cm}<{\centering}
	p{1.1cm}<{\centering} p{1.1cm}<{\centering} p{1.2cm}<{\centering}
	p{1.6cm}<{\centering} p{1cm}<{\centering}}
	\toprule
    Dataset & Metric & PopRec & FPMC & GRU4Rec & SASRec & BERT4Rec & FDSA & ZESRec & S$^{3}$-Rec & \modelname$_{t}$ & \modelname$_{t+ID}$ & \emph{Improv.} \\
	\midrule
    \multirow{2} * {Pantry}
    & HR@10 & 0.0068 & 0.0373 & 0.0395 & 0.0488 & 0.0311 & 0.0422 & \underline{0.0529} & 0.0509 & 0.0738 & \textbf{0.0750} & +41.78\% \\
    & NDCG@10 & 0.0024 & 0.0196 & 0.0194 & 0.0231 & 0.0160 & 0.0226 & \underline{0.0263} & 0.0242 & \textbf{0.0378} & 0.0375 & +43.73\% \\
    \midrule
    \multirow{2} * {Instruments}
    & HR@10 & 0.0133 & 0.1043 & 0.1045 & 0.1103 & 0.1057 & 0.1117 & 0.1076 & \underline{0.1123} &  0.1250 & \textbf{0.1304} & +16.12\% \\
    & NDCG@10 & 0.0039 & 0.0771 & 0.0796 & 0.0787 & 0.0697 & \underline{0.0840} & 0.0711 & 0.0795 & 0.0821 & \textbf{0.0872} & +3.81\% \\
    \midrule
    \multirow{2} * {Arts}
    & HR@10 & 0.0156 & 0.0958 & 0.0909 & \underline{0.1164} & 0.1096 & 0.1074 & 0.0971 & 0.1093 & 0.1130 & \textbf{0.1304} & +12.03\% \\
    & NDCG@10 & 0.0090  & 0.0684 & 0.0637 & 0.0685 & \underline{0.0774} & 0.0768 & 0.0579 & 0.0692 & 0.0708 & \textbf{0.0828} & +6.98\% \\
    \midrule
    \multirow{2} * {Food}
    & HR@10 & 0.0281  & 0.0940 & 0.1075 & \underline{0.1173} & 0.1119 & 0.1124 & 0.0967 & 0.1163  & 0.1097 & \textbf{0.1309} & +11.59\% \\
    & NDCG@10 & 0.0141 & 0.0746 & 0.0862 & 0.0846 & 0.0792 & \underline{0.0883} & 0.0646 & 0.0864 & 0.0730 & \textbf{0.0943} & +6.80\% \\
\bottomrule
\end{tabular}}}
\vspace{-0.1in}
\end{table*}

\subsubsection{Datasets.}
We conduct experiments on the Amazon review datasets~\cite{Amazon}, which contain product ratings and reviews in 29 categories on Amazon.com and rich textual metadata such as \emph{title}, \emph{brand}, \emph{description}, etc.
We use the version released in the year 2018.
Specifically, we use the 5-core data of \emph{Pantry}, \emph{Instruments}, \emph{Arts} and \emph{Food}, in which each user or item has at least 5 associated ratings.
The statistics of our datasets are summarized in Table~\ref{tab-data}.

\subsubsection{Comparison Methods.}
We consider the following baselines for comparisons:
(1) \textbf{PopRec} recommends items according to the item popularity; 
(2) \textbf{FPMC}~\cite{FPMC} models the behavior correlations by Markov chain; 
(3) \textbf{GRU4Rec}~\cite{GRU4Rec} applies GRU to model user behaviors;
(4) \textbf{SASRec}~\cite{SASRec} applies self-attention mechanism  to model user behaviors;
(5) \textbf{BERT4Rec}~\cite{BERT4Rec} applies bidirectional self-attention mechanism to model user behaviors;
(6) \textbf{FDSA}~\cite{FDSA} constructs a feature sequence and uses a feature level self-attention block to model the feature transition patterns;
(7) \textbf{ZESRec}~\cite{ZESRec} regards pre-trained BERT representations as item representations for cross-domain recommendation.
In our setting, we report its result on source domain data.
(8) \textbf{S$^3$-Rec}~\cite{S3Rec} pre-trains user behavior models via mutual information maximization objectives for feature fusion.
We implement these methods with RecBole~\cite{RecBole}.

Among all the above methods, FPMC, GRU4Rec, SASRec and BERT4Rec are general sequential recommendation methods that model the user behavior sequences only by user-item interaction data.
FDSA, ZESRec and S$^3$-Rec are text-enhanced sequential recommendation methods that model the user behavior sequences with extra information from item texts.
For our proposed approach, \textbf{\modelname$_{t}$} and \textbf{\modelname$_{t+ID}$} represent the model is fine-tuned without ID and with ID, respectively.

\subsubsection{Evaluation Settings.}
To evaluate the performance, we adopt top-$k$ Hit Ratio (HR@$k$) and top-$k$ Normalized Discounted Cumulative Gain (NDCG@$k$) evaluation metrics.
Following previous works~\cite{SASRec, BERT4Rec}, we apply the \emph{leave-one-out} strategy for evaluation.
For each user, we treat all the items that this user has not interacted with as negative items.

\subsection{Experimental Results}
\label{sec-main-exp}
In this section, we first compare the proposed IDA-SR approach with the aforementioned baselines on the four datasets, then conduct the ablation study, and finally compare the results on cold-start items.

\subsubsection{Main Results}
Compared with the general sequential recommendation methods, text-enhanced sequential recommendation methods perform better on some datasets.
Because item texts are used as auxiliary features to help improve the recommendation performance.
These results further confirm that semantic information from item text is useful for modeling user behaviors.
By comparing the proposed approach \modelname$_{t+ID}$ with all the baselines, it is clear that \modelname$_{t+ID}$ consistently performs better than them by a large margin.
Different from these baselines, we adopt the ID-agnostic user behavior pre-training framework, which transfers the semantic knowledge to guide the user behavior modeling through the three appropriate self-supervised tasks.
Then in the fine-tuning stage, we fine-tune the model to utilize the user behavior knowledge from the pre-trained model.
In this way, our proposed approach can better capture user behavior patterns and achieve much better results.
Besides, without incorporating item ID information in the fine-tuning stage, \modelname$_{t}$ also has a comparable result with other baseline methods which model user behaviors based on the ID information.
This further illustrates the effectiveness of our proposed approach.

\subsubsection{Ablation Study}
\begin{figure}[htb]
    \centering
    \subfigure[Pantry]{
        \centering
        \includegraphics[width=0.225\textwidth]{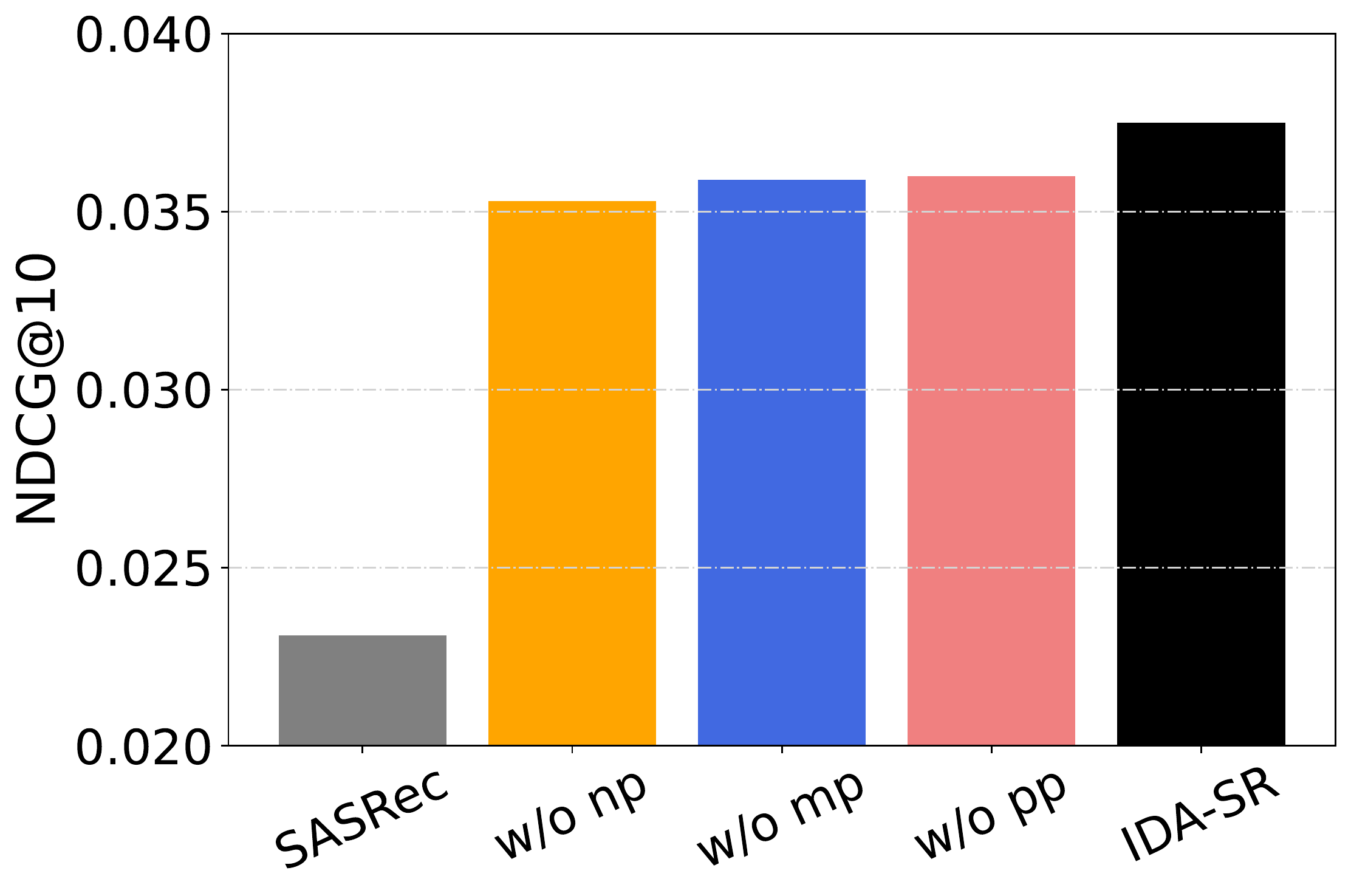}
    }
    \subfigure[Instruments]{
        \centering
        \includegraphics[width=0.225\textwidth]{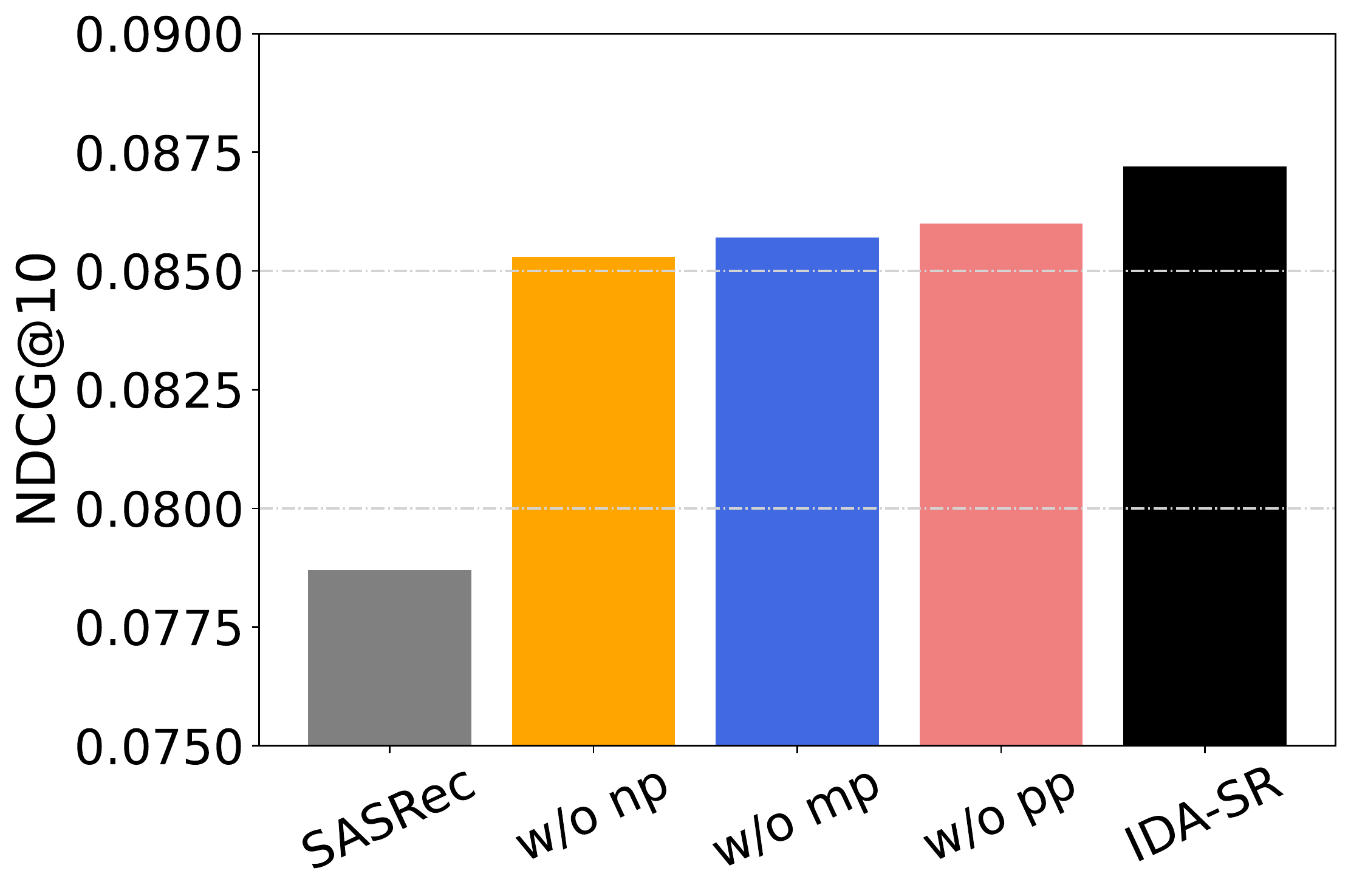}
    }
    \centering
    \caption{Ablation study.}
    \label{fig-exp-strategy}
\vspace{-0.15in}
\end{figure}

We examine the performance of IDA-SR’s
variants by removing each pre-training task from the full approach.
We use \emph{np}, \emph{mp} and \emph{pp} to represent three pre-training task \emph{next item prediction}, \emph{masked item prediction} and \emph{permuted item prediction}, respectively.
Figure~\ref{fig-exp-strategy} presents the evaluation results.
We can observe that the three pre-training tasks all contribute to the final performance.
All of them are helpful for bridging the semantic gap between text semantics and sequential preference characteristics. 

\begin{figure}[htb]
    \subfigure[Pantry]{
        \centering
        \includegraphics[width=0.225\textwidth]{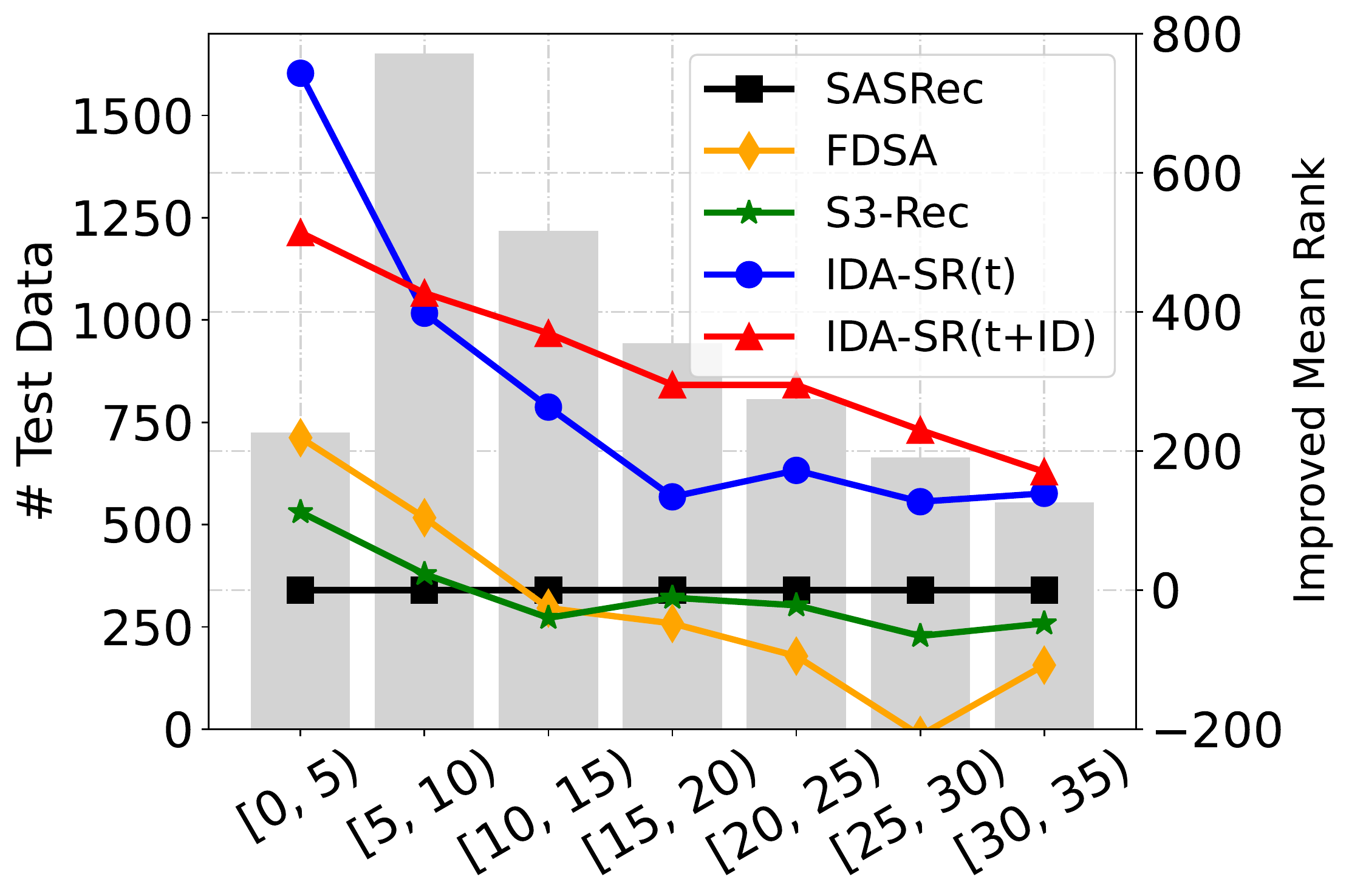}
    }
    \subfigure[Instruments]{
        \centering
        \includegraphics[width=0.225\textwidth]{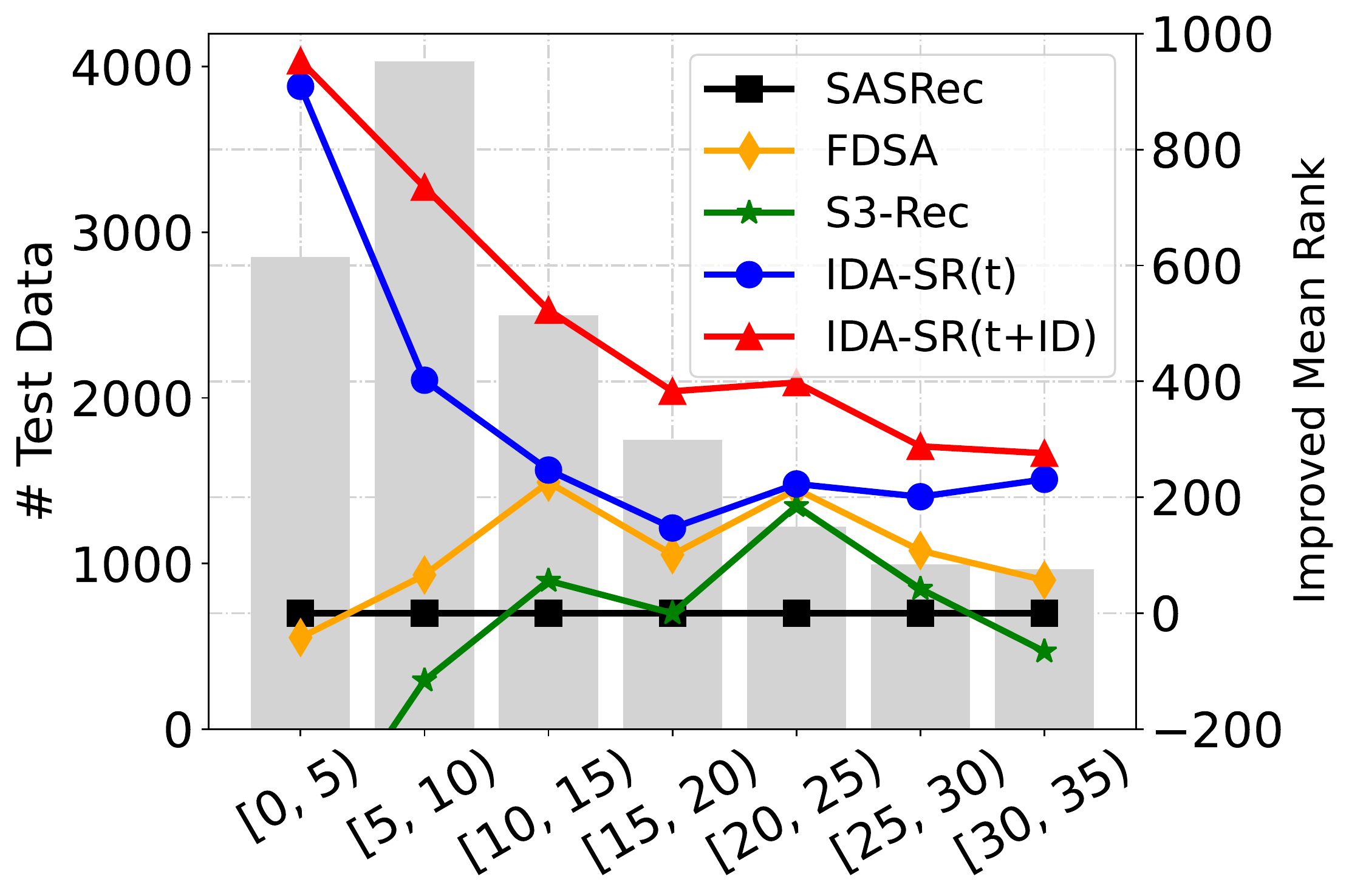}
    }
    \centering
    \caption{Performance comparison \wrt cold-start items. The bar graph represents the number of test data for each group and the line chart represents the improved mean rank of ground truth item compared with SASRec.}
    \label{fig-exp-lt}
\vspace{-0.15in}
\end{figure}

\subsubsection{Performance Comparison \wrt Cold-start Items}
Conventional user behavior modeling methods are likely to suffer from the cold-start items recommendation problem.
This problem can be alleviated by our method because the proposed ID-agnostic pre-training framework can utilize the text semantic information to make the model less dependent on the interaction data.
To verify this, we split the test data according to the popularity of ground truth items in the training data, and then record the improved mean rank of ground truth item in each group compared with baseline method SASRec.
% Figure~\ref{fig-exp-lt} shows the evaluation results on the Pantry and Instruments datasets.
From Figure~\ref{fig-exp-lt},
we can find that the proposed \modelname$_{t}$ and \modelname$_{t+ID}$ have a big improvement when the ground truth item is extremely unpopular \eg group [0, 5) and group [5, 10).
This observation implies the proposed IDA-SR can alleviate the cold-start items recommendation problem.

\section{CONCLUSION}
\label{sec-con}
In this paper, we propose the ID-agnostic user behavior modeling approach for sequential recommendation, named IDA-SR.
Different from the existing sequential recommendation methods that are limited by the ID-based item representations, the proposed IDA-SR adopts the ID-agnostic item representations based on item texts to help user behavior modeling in a direct and natural way.
To bridge the gap between text semantics and sequential user behaviors, the proposed IDA-SR conducts a pre-training architecture over item text representations on the sequential user behaviors.
Experimental results have shown the effectiveness of the proposed approach by comparing it with several competitive baselines, especially when training data is limited.
For future work, we will explore more kinds of item side information to represent items and help user behavior modeling, \eg images and videos.

\balance
\bibliographystyle{ACM-Reference-Format}
\bibliography{main}


\begin{thebibliography}{19}

%%% the Web address.

\ifx \showCODEN    \undefined \def \showCODEN     #1{\unskip}     \fi
\ifx \showDOI      \undefined \def \showDOI       #1{#1}\fi
\ifx \showISBNx    \undefined \def \showISBNx     #1{\unskip}     \fi
\ifx \showISBNxiii \undefined \def \showISBNxiii  #1{\unskip}     \fi
\ifx \showISSN     \undefined \def \showISSN      #1{\unskip}     \fi
\ifx \showLCCN     \undefined \def \showLCCN      #1{\unskip}     \fi
\ifx \shownote     \undefined \def \shownote      #1{#1}          \fi
\ifx \showarticletitle \undefined \def \showarticletitle #1{#1}   \fi
\ifx \showURL      \undefined \def \showURL       {\relax}        \fi
% invisible to TeX
\providecommand\bibfield[2]{#2}
\providecommand\bibinfo[2]{#2}
\providecommand\natexlab[1]{#1}
\providecommand\showeprint[2][]{arXiv:#2}

\bibitem[\protect\citeauthoryear{Brown, Mann, Ryder, Subbiah, Kaplan, Dhariwal,
  Neelakantan, Shyam, Sastry, Askell, Agarwal, Herbert{-}Voss, Krueger,
  Henighan, Child, Ramesh, Ziegler, Wu, Winter, Hesse, Chen, Sigler, Litwin,
  Gray, Chess, Clark, Berner, McCandlish, Radford, Sutskever, and Amodei}{Brown
  et~al\mbox{.}}{2020}]%
        {GPT3}
\bibfield{author}{\bibinfo{person}{Tom~B. Brown}, \bibinfo{person}{Benjamin
  Mann}, \bibinfo{person}{Nick Ryder}, \bibinfo{person}{Melanie Subbiah},
  \bibinfo{person}{Jared Kaplan}, \bibinfo{person}{Prafulla Dhariwal},
  \bibinfo{person}{Arvind Neelakantan}, \bibinfo{person}{Pranav Shyam},
  \bibinfo{person}{Girish Sastry}, \bibinfo{person}{Amanda Askell},
  \bibinfo{person}{Sandhini Agarwal}, \bibinfo{person}{Ariel Herbert{-}Voss},
  \bibinfo{person}{Gretchen Krueger}, \bibinfo{person}{Tom Henighan},
  \bibinfo{person}{Rewon Child}, \bibinfo{person}{Aditya Ramesh},
  \bibinfo{person}{Daniel~M. Ziegler}, \bibinfo{person}{Jeffrey Wu},
  \bibinfo{person}{Clemens Winter}, \bibinfo{person}{Christopher Hesse},
  \bibinfo{person}{Mark Chen}, \bibinfo{person}{Eric Sigler},
  \bibinfo{person}{Mateusz Litwin}, \bibinfo{person}{Scott Gray},
  \bibinfo{person}{Benjamin Chess}, \bibinfo{person}{Jack Clark},
  \bibinfo{person}{Christopher Berner}, \bibinfo{person}{Sam McCandlish},
  \bibinfo{person}{Alec Radford}, \bibinfo{person}{Ilya Sutskever}, {and}
  \bibinfo{person}{Dario Amodei}.} \bibinfo{year}{2020}\natexlab{}.
\newblock \showarticletitle{Language Models are Few-Shot Learners}. In
  \bibinfo{booktitle}{\emph{NeurIPS}}.
\newblock


\bibitem[\protect\citeauthoryear{Chen, Zhang, Liu, and Ma}{Chen
  et~al\mbox{.}}{2018}]%
        {NARRE}
\bibfield{author}{\bibinfo{person}{Chong Chen}, \bibinfo{person}{Min Zhang},
  \bibinfo{person}{Yiqun Liu}, {and} \bibinfo{person}{Shaoping Ma}.}
  \bibinfo{year}{2018}\natexlab{}.
\newblock \showarticletitle{Neural Attentional Rating Regression with
  Review-level Explanations}. In \bibinfo{booktitle}{\emph{WWW}}.
  \bibinfo{pages}{1583--1592}.
\newblock


\bibitem[\protect\citeauthoryear{Devlin, Chang, Lee, and Toutanova}{Devlin
  et~al\mbox{.}}{2019}]%
        {BERT}
\bibfield{author}{\bibinfo{person}{Jacob Devlin}, \bibinfo{person}{Ming{-}Wei
  Chang}, \bibinfo{person}{Kenton Lee}, {and} \bibinfo{person}{Kristina
  Toutanova}.} \bibinfo{year}{2019}\natexlab{}.
\newblock \showarticletitle{{BERT:} Pre-training of Deep Bidirectional
  Transformers for Language Understanding}. In
  \bibinfo{booktitle}{\emph{NAACL}}. \bibinfo{pages}{4171--4186}.
\newblock


\bibitem[\protect\citeauthoryear{Ding, Ma, Deoras, Wang, and Wang}{Ding
  et~al\mbox{.}}{2021}]%
        {ZESRec}
\bibfield{author}{\bibinfo{person}{Hao Ding}, \bibinfo{person}{Yifei Ma},
  \bibinfo{person}{Anoop Deoras}, \bibinfo{person}{Yuyang Wang}, {and}
  \bibinfo{person}{Hao Wang}.} \bibinfo{year}{2021}\natexlab{}.
\newblock \showarticletitle{Zero-Shot Recommender Systems}.
\newblock \bibinfo{journal}{\emph{arXiv preprint arXiv:2105.08318}}
  (\bibinfo{year}{2021}).
\newblock


\bibitem[\protect\citeauthoryear{Fang, Zhang, Shu, and Guo}{Fang
  et~al\mbox{.}}{2020}]%
        {sr-survey}
\bibfield{author}{\bibinfo{person}{Hui Fang}, \bibinfo{person}{Danning Zhang},
  \bibinfo{person}{Yiheng Shu}, {and} \bibinfo{person}{Guibing Guo}.}
  \bibinfo{year}{2020}\natexlab{}.
\newblock \showarticletitle{Deep Learning for Sequential Recommendation:
  Algorithms, Influential Factors, and Evaluations}.
\newblock \bibinfo{journal}{\emph{{ACM} Trans. Inf. Syst.}}
  \bibinfo{volume}{39}, \bibinfo{number}{1} (\bibinfo{year}{2020}),
  \bibinfo{pages}{10:1--10:42}.
\newblock


\bibitem[\protect\citeauthoryear{Hidasi, Karatzoglou, Baltrunas, and
  Tikk}{Hidasi et~al\mbox{.}}{2016}]%
        {GRU4Rec}
\bibfield{author}{\bibinfo{person}{Bal{\'{a}}zs Hidasi},
  \bibinfo{person}{Alexandros Karatzoglou}, \bibinfo{person}{Linas Baltrunas},
  {and} \bibinfo{person}{Domonkos Tikk}.} \bibinfo{year}{2016}\natexlab{}.
\newblock \showarticletitle{Session-based Recommendations with Recurrent Neural
  Networks}. In \bibinfo{booktitle}{\emph{ICLR}}.
\newblock


\bibitem[\protect\citeauthoryear{Kang and McAuley}{Kang and McAuley}{2018}]%
        {SASRec}
\bibfield{author}{\bibinfo{person}{Wang{-}Cheng Kang} {and}
  \bibinfo{person}{Julian~J. McAuley}.} \bibinfo{year}{2018}\natexlab{}.
\newblock \showarticletitle{Self-Attentive Sequential Recommendation}. In
  \bibinfo{booktitle}{\emph{ICDM}}. \bibinfo{pages}{197--206}.
\newblock


\bibitem[\protect\citeauthoryear{Li, Ren, Chen, Ren, Lian, and Ma}{Li
  et~al\mbox{.}}{2017}]%
        {NARM}
\bibfield{author}{\bibinfo{person}{Jing Li}, \bibinfo{person}{Pengjie Ren},
  \bibinfo{person}{Zhumin Chen}, \bibinfo{person}{Zhaochun Ren},
  \bibinfo{person}{Tao Lian}, {and} \bibinfo{person}{Jun Ma}.}
  \bibinfo{year}{2017}\natexlab{}.
\newblock \showarticletitle{Neural Attentive Session-based Recommendation}. In
  \bibinfo{booktitle}{\emph{CIKM}}. \bibinfo{pages}{1419--1428}.
\newblock


\bibitem[\protect\citeauthoryear{Liu, Fan, Wang, and Yu}{Liu
  et~al\mbox{.}}{2021}]%
        {ASRep}
\bibfield{author}{\bibinfo{person}{Zhiwei Liu}, \bibinfo{person}{Ziwei Fan},
  \bibinfo{person}{Yu Wang}, {and} \bibinfo{person}{Philip~S. Yu}.}
  \bibinfo{year}{2021}\natexlab{}.
\newblock \showarticletitle{Augmenting Sequential Recommendation with
  Pseudo-Prior Items via Reversely Pre-training Transformer}. In
  \bibinfo{booktitle}{\emph{SIGIR}}. \bibinfo{pages}{1608--1612}.
\newblock


\bibitem[\protect\citeauthoryear{Ni, Li, and McAuley}{Ni et~al\mbox{.}}{2019}]%
        {Amazon}
\bibfield{author}{\bibinfo{person}{Jianmo Ni}, \bibinfo{person}{Jiacheng Li},
  {and} \bibinfo{person}{Julian~J. McAuley}.} \bibinfo{year}{2019}\natexlab{}.
\newblock \showarticletitle{Justifying Recommendations using Distantly-Labeled
  Reviews and Fine-Grained Aspects}. In \bibinfo{booktitle}{\emph{EMNLP}}.
  \bibinfo{pages}{188--197}.
\newblock


\bibitem[\protect\citeauthoryear{Rendle, Freudenthaler, and
  Schmidt{-}Thieme}{Rendle et~al\mbox{.}}{2010}]%
        {FPMC}
\bibfield{author}{\bibinfo{person}{Steffen Rendle}, \bibinfo{person}{Christoph
  Freudenthaler}, {and} \bibinfo{person}{Lars Schmidt{-}Thieme}.}
  \bibinfo{year}{2010}\natexlab{}.
\newblock \showarticletitle{Factorizing personalized Markov chains for
  next-basket recommendation}. In \bibinfo{booktitle}{\emph{WWW}}.
  \bibinfo{pages}{811--820}.
\newblock


\bibitem[\protect\citeauthoryear{Sun, Liu, Wu, Pei, Lin, Ou, and Jiang}{Sun
  et~al\mbox{.}}{2019}]%
        {BERT4Rec}
\bibfield{author}{\bibinfo{person}{Fei Sun}, \bibinfo{person}{Jun Liu},
  \bibinfo{person}{Jian Wu}, \bibinfo{person}{Changhua Pei},
  \bibinfo{person}{Xiao Lin}, \bibinfo{person}{Wenwu Ou}, {and}
  \bibinfo{person}{Peng Jiang}.} \bibinfo{year}{2019}\natexlab{}.
\newblock \showarticletitle{BERT4Rec: Sequential Recommendation with
  Bidirectional Encoder Representations from Transformer}. In
  \bibinfo{booktitle}{\emph{CIKM}}. \bibinfo{pages}{1441--1450}.
\newblock


\bibitem[\protect\citeauthoryear{Tang and Wang}{Tang and Wang}{2018}]%
        {Caser}
\bibfield{author}{\bibinfo{person}{Jiaxi Tang} {and} \bibinfo{person}{Ke
  Wang}.} \bibinfo{year}{2018}\natexlab{}.
\newblock \showarticletitle{Personalized Top-N Sequential Recommendation via
  Convolutional Sequence Embedding}. In \bibinfo{booktitle}{\emph{WSDM}}.
  \bibinfo{pages}{565--573}.
\newblock


\bibitem[\protect\citeauthoryear{Vaswani, Shazeer, Parmar, Uszkoreit, Jones,
  Gomez, Kaiser, and Polosukhin}{Vaswani et~al\mbox{.}}{2017}]%
        {Trans}
\bibfield{author}{\bibinfo{person}{Ashish Vaswani}, \bibinfo{person}{Noam
  Shazeer}, \bibinfo{person}{Niki Parmar}, \bibinfo{person}{Jakob Uszkoreit},
  \bibinfo{person}{Llion Jones}, \bibinfo{person}{Aidan~N. Gomez},
  \bibinfo{person}{Lukasz Kaiser}, {and} \bibinfo{person}{Illia Polosukhin}.}
  \bibinfo{year}{2017}\natexlab{}.
\newblock \showarticletitle{Attention is All you Need}. In
  \bibinfo{booktitle}{\emph{{NIPS}}}. \bibinfo{pages}{5998--6008}.
\newblock


\bibitem[\protect\citeauthoryear{Yang, Dai, Yang, Carbonell, Salakhutdinov, and
  Le}{Yang et~al\mbox{.}}{2019}]%
        {XLNet}
\bibfield{author}{\bibinfo{person}{Zhilin Yang}, \bibinfo{person}{Zihang Dai},
  \bibinfo{person}{Yiming Yang}, \bibinfo{person}{Jaime~G. Carbonell},
  \bibinfo{person}{Ruslan Salakhutdinov}, {and} \bibinfo{person}{Quoc~V. Le}.}
  \bibinfo{year}{2019}\natexlab{}.
\newblock \showarticletitle{XLNet: Generalized Autoregressive Pretraining for
  Language Understanding}. In \bibinfo{booktitle}{\emph{NeurIPS}}.
\newblock


\bibitem[\protect\citeauthoryear{Yu, Lin, Ge, Ou, and Qin}{Yu
  et~al\mbox{.}}{2020}]%
        {TCF}
\bibfield{author}{\bibinfo{person}{Wenhui Yu}, \bibinfo{person}{Xiao Lin},
  \bibinfo{person}{Junfeng Ge}, \bibinfo{person}{Wenwu Ou}, {and}
  \bibinfo{person}{Zheng Qin}.} \bibinfo{year}{2020}\natexlab{}.
\newblock \showarticletitle{Semi-supervised Collaborative Filtering by
  Text-enhanced Domain Adaptation}. In \bibinfo{booktitle}{\emph{SIGKDD}}.
  \bibinfo{pages}{2136--2144}.
\newblock


\bibitem[\protect\citeauthoryear{Zhang, Zhao, Liu, Sheng, Xu, Wang, Liu, and
  Zhou}{Zhang et~al\mbox{.}}{2019}]%
        {FDSA}
\bibfield{author}{\bibinfo{person}{Tingting Zhang}, \bibinfo{person}{Pengpeng
  Zhao}, \bibinfo{person}{Yanchi Liu}, \bibinfo{person}{Victor~S. Sheng},
  \bibinfo{person}{Jiajie Xu}, \bibinfo{person}{Deqing Wang},
  \bibinfo{person}{Guanfeng Liu}, {and} \bibinfo{person}{Xiaofang Zhou}.}
  \bibinfo{year}{2019}\natexlab{}.
\newblock \showarticletitle{Feature-level Deeper Self-Attention Network for
  Sequential Recommendation}. In \bibinfo{booktitle}{\emph{IJCAI}}.
  \bibinfo{pages}{4320--4326}.
\newblock


\bibitem[\protect\citeauthoryear{Zhao, Mu, Hou, Lin, Chen, Pan, Li, Lu, Wang,
  Tian, Min, Feng, Fan, Chen, Wang, Ji, Li, Wang, and Wen}{Zhao
  et~al\mbox{.}}{2021}]%
        {RecBole}
\bibfield{author}{\bibinfo{person}{Wayne~Xin Zhao}, \bibinfo{person}{Shanlei
  Mu}, \bibinfo{person}{Yupeng Hou}, \bibinfo{person}{Zihan Lin},
  \bibinfo{person}{Yushuo Chen}, \bibinfo{person}{Xingyu Pan},
  \bibinfo{person}{Kaiyuan Li}, \bibinfo{person}{Yujie Lu},
  \bibinfo{person}{Hui Wang}, \bibinfo{person}{Changxin Tian},
  \bibinfo{person}{Yingqian Min}, \bibinfo{person}{Zhichao Feng},
  \bibinfo{person}{Xinyan Fan}, \bibinfo{person}{Xu Chen},
  \bibinfo{person}{Pengfei Wang}, \bibinfo{person}{Wendi Ji},
  \bibinfo{person}{Yaliang Li}, \bibinfo{person}{Xiaoling Wang}, {and}
  \bibinfo{person}{Ji{-}Rong Wen}.} \bibinfo{year}{2021}\natexlab{}.
\newblock \showarticletitle{RecBole: Towards a Unified, Comprehensive and
  Efficient Framework for Recommendation Algorithms}. In
  \bibinfo{booktitle}{\emph{CIKM}}. \bibinfo{pages}{4653--4664}.
\newblock


\bibitem[\protect\citeauthoryear{Zhou, Wang, Zhao, Zhu, Wang, Zhang, Wang, and
  Wen}{Zhou et~al\mbox{.}}{2020}]%
        {S3Rec}
\bibfield{author}{\bibinfo{person}{Kun Zhou}, \bibinfo{person}{Hui Wang},
  \bibinfo{person}{Wayne~Xin Zhao}, \bibinfo{person}{Yutao Zhu},
  \bibinfo{person}{Sirui Wang}, \bibinfo{person}{Fuzheng Zhang},
  \bibinfo{person}{Zhongyuan Wang}, {and} \bibinfo{person}{Ji{-}Rong Wen}.}
  \bibinfo{year}{2020}\natexlab{}.
\newblock \showarticletitle{S3-Rec: Self-Supervised Learning for Sequential
  Recommendation with Mutual Information Maximization}. In
  \bibinfo{booktitle}{\emph{CIKM}}. \bibinfo{pages}{1893--1902}.
\newblock


\end{thebibliography}

\end{document}